\documentclass[aps,twocolumn,nofootinbib,showkeys,showpacs,superscriptaddress]{revtex4-1}
\usepackage{amssymb}
\usepackage{amsmath}
\usepackage{amstext}
\usepackage{amsfonts}
\usepackage{bbold}
\usepackage{bm}
\usepackage{graphics}
\usepackage{mathtools}
\usepackage{lmodern}
\usepackage{graphicx}
\usepackage{hyperref}
\usepackage{xcolor}

\begin{document}

\title{Sub-MeV Dark Matter and the Goldstone Modes of  Superfluid Helium} 

\author{Andrea~Caputo}
\affiliation{Instituto de Fisica Corpuscular, Universidad de Valencia and CSIC, Edificio Institutos Investigacion, Catedratico Jose Beltran 2, Paterna, 46980 Spain}

\author{Angelo~Esposito}
\affiliation{Theoretical Particle Physics Laboratory (LPTP), Institute of Physics, EPFL, 1015 Lausanne, Switzerland}

\author{Antonio~D.~Polosa}
\affiliation{Dipartimento di Fisica and INFN, Sapienza Universit\`a di Roma, P.le Aldo Moro 2, I-00185 Roma, Italy}

\begin{abstract}
We show how a relativistic effective field theory for the superfluid phase of $^4$He can replace the standard methods used to compute the production rates of low momentum excitations due to the interaction with an external probe. This is done by studying the scattering problem of a light dark matter particle in the superfluid, and comparing to some existing results.
We show that the rate of emission of two phonons, the Goldstone modes of the effective theory,  gets strongly suppressed  for sub-MeV dark matter particles due to a fine cancellation between two different tree-level diagrams in the limit of small exchanged momenta. This phenomenon is found to be a consequence of  the particular choice of the potential felt by the dark matter particle  in helium. The predicted rates can vary by orders of magnitude if this potential is changed.
We prove that the dominant contribution to the total emission rate is provided by excitations in the phonon branch.
Finally, we analyze the angular distributions for the emissions of one and two phonons, and discuss how they can be used to measure the mass of the hypothetical dark matter particle hitting the helium target.
\end{abstract}


\maketitle


\section{Introduction}

The presence of dark matter is one of the most compelling evidences for physics beyond the Standard Model, but the question about its nature remains unanswered. Following the negative results of the searches for the so-called Weakly Interacting Massive Particles, more attention is being given to the study of models of dark matter with masses below the GeV, whose search requires new ideas and  detection techniques~\cite{Essig:2011nj,Graham:2012su,Capparelli:2014lua,Essig:2015cda,Hochberg:2015pha,Hochberg:2015fth,Cavoto:2016lqo,Hochberg:2016ntt,Cavoto:2017otc, Hochberg:2017wce, Fichet:2017bng, Alonso:2018dxy,Trickle:2019ovy, Cheng:2019vwy,Dror:2019onn} | for a review see e.g.~\cite{Alexander:2016aln,Battaglieri:2017aum, Lin:2019uvt}.

A promising proposal is that of employing superfluid $^4$He, whose properties are particularly amenable for the search of dark matter as light as the keV~\cite{Schutz:2016tid,Knapen:2016cue}. This is especially true because of the presence of gapless excitations. The newly developed effective field theory (EFT) for superfluids~\cite{Leutwyler:1996er,Son:2002zn,Nicolis:2011cs,Nicolis:2015sra} has been recently  employed to describe their interaction with the dark matter particles, with particular focus on masses between the GeV and the MeV~\cite{Acanfora:2019con}.

In this paper we confirm the results of our previous analysis~\cite{Acanfora:2019con}, and study the case of dark matter particles with masses as low as few keVs. This is the region where a detector based on $^4$He is most competitive, and the physics behind the emission of two phonons, induced by the passage of the dark matter,  presents non-trivial aspects, which were not appreciated so far. We show that if the in-medium potential for the dark matter is proportional to the $^4$He density, then the rate of emission of two phonons is highly suppressed with respect to what expected on the basis of phase space considerations only\footnote{A somewhat similar effect is also observed in crystals for the emission of a single optical phonon, and it would be interesting to understand whether any relation exists~\cite{Cox:2019cod}.}. Our results agree very well with those found in~\cite{Schutz:2016tid,Knapen:2016cue}, but from the EFT we are able to understand that the predicted  suppression is due to a precise cancellation between two tree-level diagrams~\cite{Acanfora:2019con}, which occurs in the limit of small momentum transfers. This turns out to be the consequence of integrating out highly off-shell phonons. 

We observe that this mechanism does not take place if the interaction between the dark matter and the bulk of $^4$He happens via a coupling different from the simple number density. In this case the emission rate can be dramatically increased, leading to much stronger exclusion limits. 

Comparing with the results obtained in~\cite{Schutz:2016tid,Knapen:2016cue}, we also deduce that most of the contribution to the emission rate is due to final state phonons, with very little role played by higher momentum excitations like maxons and rotons.

Finally, we show how the angular distributions of the final state  phonons encode  information about the dark matter mass, and could  be used to measure it. This is true for both the single and two phonon emissions.

\vspace{0.5em}

\noindent\emph{Conventions:} Throughout this paper we set $\hbar = c = 1$ and work with a metric signature $\eta_{\mu\nu}=\text{diag}(-1,1,1,1)$.

\section{Effective action and two-phonon emission}

From the EFT viewpoint a superfluid like $^4$He is a system characterized by a $U(1)$ symmetry (particle number), whose charge $Q$ is at finite density and is spontaneously broken. On top of that, a superfluid also breaks boosts and time translations, but preserves a combination of them~\cite{Nicolis:2011pv}. This symmetry breaking pattern can be implemented via a real scalar field $\psi(x)$ which shifts under the $U(1)$, $\psi\to\psi+a$, and acquires a vacuum expectation value proportional to time, $\langle \psi(x)\rangle = \mu t$. Its fluctuation corresponds to the superfluid Goldstone mode, the phonon $\pi$ in  $\psi(x)=\mu t+c_s\,\sqrt{\mu/\bar n}\,\pi(x)$, where $\bar n$, $c_s$ and $\mu$ are respectively the equilibrium number density, the speed of sound and the relativistic chemical potential of the superfluid\footnote{In the relativistic case the chemical potential gets a contribution from the rest mass of the superfluid's constituents, $\mu = m_\text{He}+\mu_\text{nr}$, with $\mu_\text{nr}$ being the nonrelativistic chemical potential.}. The coefficient has been chosen so that the phonon field is canonically normalized | see Eq.\eqref{eq:Seff}.

In~\cite{Acanfora:2019con} it was shown that, if the dark matter $\chi$ is a complex scalar charged under a new $U_d(1)$ group, and it couples to the superfluid number density through a massive scalar mediator,  the interactions of interest read
\begin{align} \label{eq:Seff}
S_\text{eff}&=\int d^4x\bigg[\frac{1}{2}\dot\pi^2-\frac{c_s^2}{2}(\bm \nabla \pi)^2 -|\partial \chi|-m_\chi^2|\chi|^2 \notag \\
&+\lambda_3\sqrt{\frac{m_\text{He}}{\bar n}}c_s\dot\pi(\bm \nabla \pi)^2+\lambda_3^\prime\sqrt{\frac{m_\text{He}}{\bar n}}c_s\dot\pi^3 \\
&-\bigg(g_1\sqrt{\frac{m_\text{He}}{\bar n}}c_s\dot\pi-\frac{g_1}{2}\frac{c_s^2}{\bar n}(\bm \nabla\pi)^2+\frac{g_2}{2}\frac{m_\text{He} c_s^2}{\bar n}\dot\pi^2\bigg)|\chi|^2\bigg] \notag.
\end{align}
The second line describes the phonon self-interactions, and the last one its interactions with the dark matter. Being this a low-energy EFT, it can only  describe excitations at low momenta, $q\lesssim \Lambda=1$~keV, {\it i.e.} the phonons. Other excitations like rotons and maxons cannot be included in this scheme. From now on we then require that the momenta of on-shell phonons never exceed $\Lambda$, and their energies never exceed $c_s\Lambda$. 

All the effective couplings are determined solely by the superfluid equation of state | in the form $P=P(\mu)$ or $c_s=c_s(P)$ | which is experimentally known~\cite{abraham1970velocity}. No input from the microscopic strongly coupled theory is needed. In particular, in the nonrelativistic limit one finds~\cite{Acanfora:2019con}
\begin{align} \label{eq:couplings}
\begin{split}
&\qquad\qquad c_s^2=\frac{1}{m_\text{He}}\frac{dP}{d\bar n}\,,\qquad\lambda_3=-\frac{1}{2m_\text{He}}\,, \\
&\lambda_3^\prime=\frac{1}{6m_\text{He}c_s^2}-\frac{\bar n}{3c_s}\frac{dc_s}{dP}\,, \qquad g_1=-G_\chi m_\chi \frac{\bar n}{m_\text{He}c_s^2}\,,\\ 
& \qquad \quad g_2=-G_\chi m_\chi \frac{\bar n}{m_\text{He}c_s^2}\left(\frac{1}{m_\text{He}c_s^2}-\frac{2\bar n}{c_s}\frac{dc_s}{dP}\right)\,,
\end{split}
\end{align}
where $P$ is the pressure of the superfluid, and $G_\chi$ an effective dark matter-helium coupling with dimension $(\text{mass})^{-2}$. 
The helium number density is $\bar n\simeq8.5 \times 10^{22}$~cm$^{-3}$, while the sound speed and its derivative are $c_s\simeq248$~m/s and $dc_s/dP\simeq8$~m/s/atm~\cite{abraham1970velocity}.

\begin{figure}[t]
\centering
\includegraphics[width=0.4\textwidth]{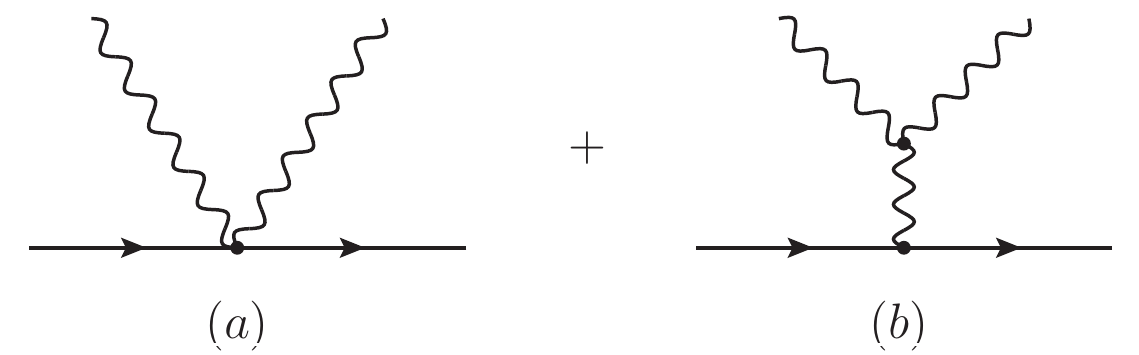}
\caption{Leading order diagrams for the emission of two phonons by the dark matter.} \label{fig:Feynman}
\end{figure}

\vspace{1em}

The amplitude for the emission of two phonons with energies and momenta $(\omega_1,\bm q_1)$ and $(\omega_2,\bm q_2)$ is obtained from the two diagrams in Figure~\ref{fig:Feynman}, whose matrix elements are~\cite{Acanfora:2019con}
\begin{subequations}
\begin{align}
\mathcal{M}_a&\!=\frac{c_s^2}{\bar n}\left(m_\text{He}g_2\omega_1\omega_2-g_1\bm q_1\cdot\bm q_2\right)\,, \\
\mathcal{M}_b&\!=-2g_1\frac{m_\text{He}c_s^2}{\bar n}\frac{\omega}{\omega^2-c_s^2\bm q^2}  \\
&\!\times\!\bigg[ \lambda_3(\omega_1\bm q_2\cdot\bm q+\omega_2\bm q_1\cdot\bm q+\omega\bm q_1\cdot\bm q_2)+3\lambda_3^\prime\omega_1\omega_2\omega\bigg]\,. \notag
\end{align}
\end{subequations}
Here $\omega=\omega_1+\omega_2$ and $\bm q=\bm q_1+\bm q_2$ are respectively the energy and momentum transferred. Note that neither $\mathcal{M}_a$ nor the last two terms of $\mathcal{M}_b$ are present in the amplitude of~\cite{Schutz:2016tid}, while they seem to be included in~\cite{Knapen:2016cue}\footnote{We are grateful to T.~Lin for pointing this out.}.  Given the different approach followed in~\cite{Schutz:2016tid,Knapen:2016cue}, a comparison in terms of diagrams is not straightforward.
The presence of  diagram $(a)$ is manifest in  the EFT, and the additional terms in $\mathcal{M}_b$ are necessary to ensure that the three-phonon vertex is invariant under the exchange of all momenta, as required by Bose symmetry.

The total rate for the two-phonon emission can be computed as in~\cite{Acanfora:2019con}, {\it i.e.}
\begin{align} \label{eq:gamma}
\Gamma=\frac{1}{8(2\pi)^4c_s^5m_\chi^2v_\chi}\int_\mathcal{R}d\theta_{12}d\theta_2d\omega_1d\omega_2\,\omega_2\frac{\big|\mathcal{M}\big|^2}{\sqrt{1-\mathcal{A}^2}}\,,
\end{align}
where $\theta_{12}$ is the angle between the two outgoing phonons, $\theta_2$ the angle of one of them with respect to the direction of the incoming dark matter, and $v_\chi$ the dark matter velocity. The function $\mathcal{A}$ is given by
\begin{align} \label{eq:A}
\begin{split}
&\mathcal{A}(\theta_{12},\theta_2,\omega_1,\omega_2)= \frac{1}{\sin\theta_{12}\sin\theta_2}\Big(\cos\theta_{12}\cos\theta_2 \\ 
&\qquad+\frac{\omega_2}{\omega_1}\cos\theta_2-\frac{\omega_2}{c_s P}\cos\theta_{12}-\frac{\omega_1^2+\omega_2^2}{2\omega_1 c_s P}\,\Big)\,.
\end{split}
\end{align}
Moreover, the integration region $\mathcal{R}$ is defined by the conditions $-1\leq\mathcal{A}\leq1$ and $k_i-k_f\leq q \leq k_i+k_f$, where $k_i=m_\chi v_\chi$ and $k_f=\sqrt{m_\chi^2v_\chi^2-2 m_\chi \omega}$ are the initial and final dark matter momenta. 

The outgoing phonons can in principle be detected with two techniques. The first is a calorimetric one, which can be used if the net energy released is $\omega\geq1$~meV~\cite{Hertel:2018aal}. The second is via the so-called quantum evaporation~\cite{Maris:2017xvi,Hertel:2018aal}, which might allow to detect single phonons but requires that the energy of each of them is larger than the surface binding energy of $^4$He, $\omega_i\geq0.62$~meV\footnote{In this range of energies the phonon is stable against the decay into two other phonons~\cite{Maris:1977zz,McKinsey}.}.

One should in principle convolute Eq.~\eqref{eq:gamma} with a Maxwell-Boltzmann distribution for the dark matter velocity but, for the sake of our arguments, it is enough to just fix the velocity to its most probable value, $v_\chi=220$~km/s. The rate of events per unit target mass is then $R=\frac{\rho_\chi}{m_\text{He}\bar n m_\chi}\Gamma$, with $\rho_\chi=0.3$~GeV/cm$^3$ being the dark matter mass density~\cite{Bovy:2012tw}.

Finally, the dark matter-proton cross section can be expressed in terms of the effective coupling $G_\chi$ using~\cite{Acanfora:2019con}
\begin{align}
\sigma_p\sim\frac{G_\chi^2}{16\pi A^2}\frac{m_\chi^2m_\text{He}^2}{{(m_\chi+m_\text{He})}^2}\,,
\end{align}
where $A=4$ for $^4$He.

\vspace{1em}

As argued in~\cite{Schutz:2016tid,Knapen:2016cue}, and as can be seen explicitly from Eq.~\eqref{eq:A}, when the mass of the dark matter particle  is brought down to around few keVs, the most relevant kinematical configuration is the one where the two outgoing phonons are almost back-to-back, $\omega_1\simeq\omega_2$ and $\bm q_1\simeq -\bm q_2$. In this limit $\mathcal{M}_a$ and the last two terms of $\mathcal{M}_b$ are precisely the dominant ones, and one might wonder what is the reason for the suppression of the emission rate observed in~\cite{Schutz:2016tid,Knapen:2016cue}. Indeed, in this limit, 
\begin{align} \mathcal{M}_a+\mathcal{M}_b=O(q^2/\omega^2)\end{align} because of an \emph{exact} cancellation between the leading order terms in the two matrix elements, intimately related to the expressions of the couplings~\eqref{eq:couplings} in terms of the equation of state.

This cancellation can be understood as the consequence of integrating out highly off-shell phonons, for which $\omega\gg c_s q$. Indeed, integrating out the off-shell phonon from diagram $(b)$ in the $\bm q=0$ limit, produces a diagram which exactly cancels the first one. To show that, let us separate the phonon field in two components, $\pi=\pi_0+\pi_1$, where $\pi_1$ only has support on a region with $\omega\gg c_sq$, and $\pi_0$ everywhere else. In particular, $\pi_0$ contains all on-shell phonons\footnote{This procedure is similar to what is typically done in NRQCD~\cite{RothTasi}.}. For highly off-shell phonons we can neglect $\bm \nabla\pi_1$. Plugging $\pi=\pi_0+\pi_1$ into the action~\eqref{eq:Seff} and working to quadratic order in $\pi_1$, produces the following terms in the nonrelativistic limit
\begin{align} \label{eq:Spi1}
S_\text{eff}&\supset\int d^4x\bigg[ \frac{1}{2}\dot \pi_1^2+\lambda_3\sqrt{\frac{m_\text{He}}{\bar n}}c_s\dot\pi_1\big(\bm \nabla\pi_0\big)^2 \notag \\
&\quad+3\lambda_3^\prime \sqrt{\frac{m_\text{He}}{\bar n}}c_s\big( \dot\pi_0^2\dot\pi_1+\dot\pi_0\dot\pi_1^2 \big)-g_1\sqrt{\frac{m_\text{He}}{\bar n}}c_s\dot\pi_1|\chi|^2 \notag \\
&\quad-\frac{g_2}{2}\frac{m_\text{He}c_s^2}{\bar n}\big(2\dot\pi_0\dot\pi_1+\dot\pi_1^2\big)|\chi|^2 \bigg]\,.
\end{align}
This describes the coupling between the dark matter, the on-shell phonon and the off-shell one.
One can now integrate out the latter at tree level. This amounts to solve its equation of motion and plug it back into the previous action. Such a solution, at lowest order in $G_\chi$ and small fluctuations, is given by
\begin{align}
\begin{split}
\dot\pi_1&=-\lambda_3\sqrt{\frac{m_\text{He}}{\bar n}}c_s\big(\bm \nabla\pi_0\big)^2-3\lambda_3^\prime\sqrt{\frac{m_\text{He}}{\bar n}}c_s\dot \pi_0^2 \\
&\quad + g_1\sqrt{\frac{m_\text{He}}{\bar n}}c_s|\chi|^2+g_2\frac{m_\text{He}c_s^2}{\bar n}\dot\pi_0|\chi|^2 \\
&\quad-6g_1\lambda_3^\prime\frac{m_\text{He}c_s^2}{\bar n}\dot\pi_0|\chi|^2+O\big(\pi_0^3,\pi_0^2|\chi|^2\big)\,.
\end{split}
\end{align}
Plugging this back into the action~\eqref{eq:Spi1} (namely in the second line and in the first term of the third), and recalling the expressions~\eqref{eq:couplings}, one can see that the resulting effective operator cancels precisely the last two terms in Eq.~\eqref{eq:Seff}. This means that the dark matter coupling to two phonons is zero in the exact $\bm q=0$. It will then be of order $q^2/\omega^2$, which is suppressed in the limit of small momentum transfer.

\begin{figure}[t]
\centering
\includegraphics[width=0.47\textwidth]{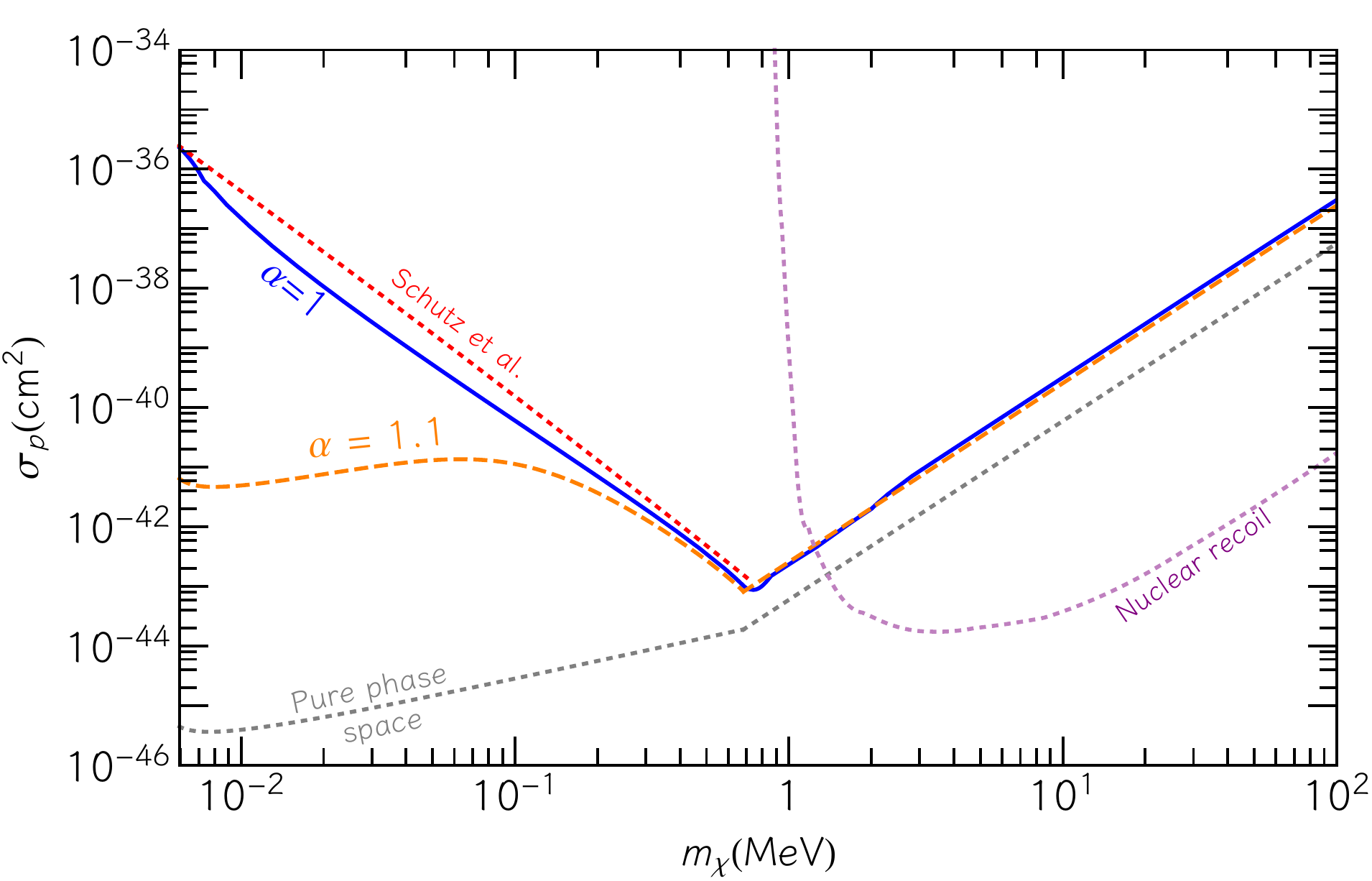}
\caption{Excluded region as referred to the $95\%$ C.L., corresponding to 3 events/kg/year, assuming zero background. We also assumed that the phonons detection happens via calorimetric techniques, imposing that the total energy released is at least $1$ meV. At masses lighter than 1~MeV, the curve obtained requiring quantum evaporation is suppressed with respect to the present one by only a small factor of order unity. The blue solid line corresponds to a model where the dark matter in-medium potential is proportional to the helium number density, the orange dashed line is instead a model where this couplings is slightly modified (see Eq.~\eqref{eq:alphamodel}). The dotted gray curve corresponds to a pure phase space behavior, obtained setting $\mathcal{M}=G_\chi m_\chi m_\text{He} c_s^2$ for dimensional reasons. The red dotted line is the one obtained in~\cite{Schutz:2016tid} with traditional methods, reported here for comparison. Finally, note that for masses heavier than 1 MeV, the dominant process is the nuclear recoil (purple dotted line)~\cite{Schutz:2016tid}.} \label{fig:exclusion}
\end{figure}

\vspace{1em}

As anticipated above, this cancellation causes the total rate to be very suppressed with respect to what expected on the basis of phase space arguments only, as one can see in Figure~\ref{fig:exclusion}. We observe that this suppression is only effective when the coupling between the dark matter and the superfluid phonon is exactly proportional to the superfluid density. To further show that, we consider an artificial model in which the dark matter-phonon coupling is
\begin{align} \label{eq:alphamodel}
S_\text{eff}\supset\int d^4x \,G_\chi m_\chi |\chi|^2\,n(X)^\alpha \,\bar n^{\,1-\alpha}\,.
\end{align}
When $\alpha=1$ one recovers the in-medium interaction used in~\cite{Schutz:2016tid,Knapen:2016cue,Acanfora:2019con}. Here $n(X)$ is the number density as a function of the local chemical potential, related to the phonon field by~\cite{Son:2002zn,Nicolis:2011cs,Acanfora:2019con}
\begin{align}
X=\mu + \sqrt{\frac{\mu}{\bar n}}c_s\dot\pi-\frac{c_s^2}{2\bar n}\big(\bm \nabla \pi\big)^2+\dots\,.
\end{align}
Note that this model is not meant to be a realistic one. It should be regarded as a toy example where the leading order cancellation explained above does not happen. 

With the interaction in~\eqref{eq:alphamodel} one gets $g_1= - G_\chi m_\chi \alpha \,\bar n^\prime$ and $g_2=-G_\chi m_\chi \alpha\big((\alpha-1)(\bar n^\prime)^2+\bar n \, \bar n^{\prime\prime}\big)/\bar n$, to be compared with those given in~\eqref{eq:couplings}. In Figure~\ref{fig:exclusion} we show that even a small deviation from a potential exactly proportional to the number density causes a drastic increase in the total rate and, consequently, a much more constraining projection. It is worth investigating to which extent different models of dark matter might predict different couplings with the bulk of $^4$He.

\vspace{1em}

Another comment is in order. We notice that the excluded region computed here for the $\alpha=1$ case agrees very well with the one reported in~\cite{Schutz:2016tid,Knapen:2016cue}, obtained with the standard treatment of superfluid helium and tested on neutron scattering data. Given that our EFT only accounts for low momentum excitations, whereas the standard approach can get to much higher momenta, we conclude that the main contribution to the total rate is due to the phonons. Other collective excitations like maxons or rotons play only a marginal role.

\vspace{1em}

Finally, we find that for masses below 1~MeV the excluded region obtained assuming that both phonons induce quantum evaporation is only marginally suppressed with respect to the one presented in Figure~\ref{fig:exclusion}, yielding essentially the same result.

\section{Dark matter mass from angular distributions}

We now show that the angular distributions of the final state phonons encode important information about the mass of the dark matter candidate, and could be used to measure it.

To reconstruct such distributions one must be able to detect single phonons, and we therefore assume that the detection happens via quantum evaporation.

Before studying the angular distributions for the two-phonon emission analyzed previously, let us discuss the case of a a single phonon emission~\cite{Acanfora:2019con}. Given the constraint on the minimum energy of a detectable phonon, this process turns out to be only effective for dark matter particles heavier than 1 MeV. Nonetheless, when allowed, its rate is the dominant one, and the angle of the outgoing phonon with respect to the incoming dark matter is completely fixed by kinematics. In particular, it is given by the \v Cherenkov angle:
\begin{align}
\cos\theta=\frac{c_s}{v_\chi}+\frac{q}{2m_\chi v_\chi}\,,
\end{align}
with $q$ being the phonon's momentum. The latter can be deduced from the energy of the evaporated helium atom, while the dark matter velocity and direction are roughly known. It then follows that a measurement of the emission angle could allow to estimate the mass of the dark matter candidate.

\begin{figure}[t]
\centering
\includegraphics[width=0.47\textwidth]{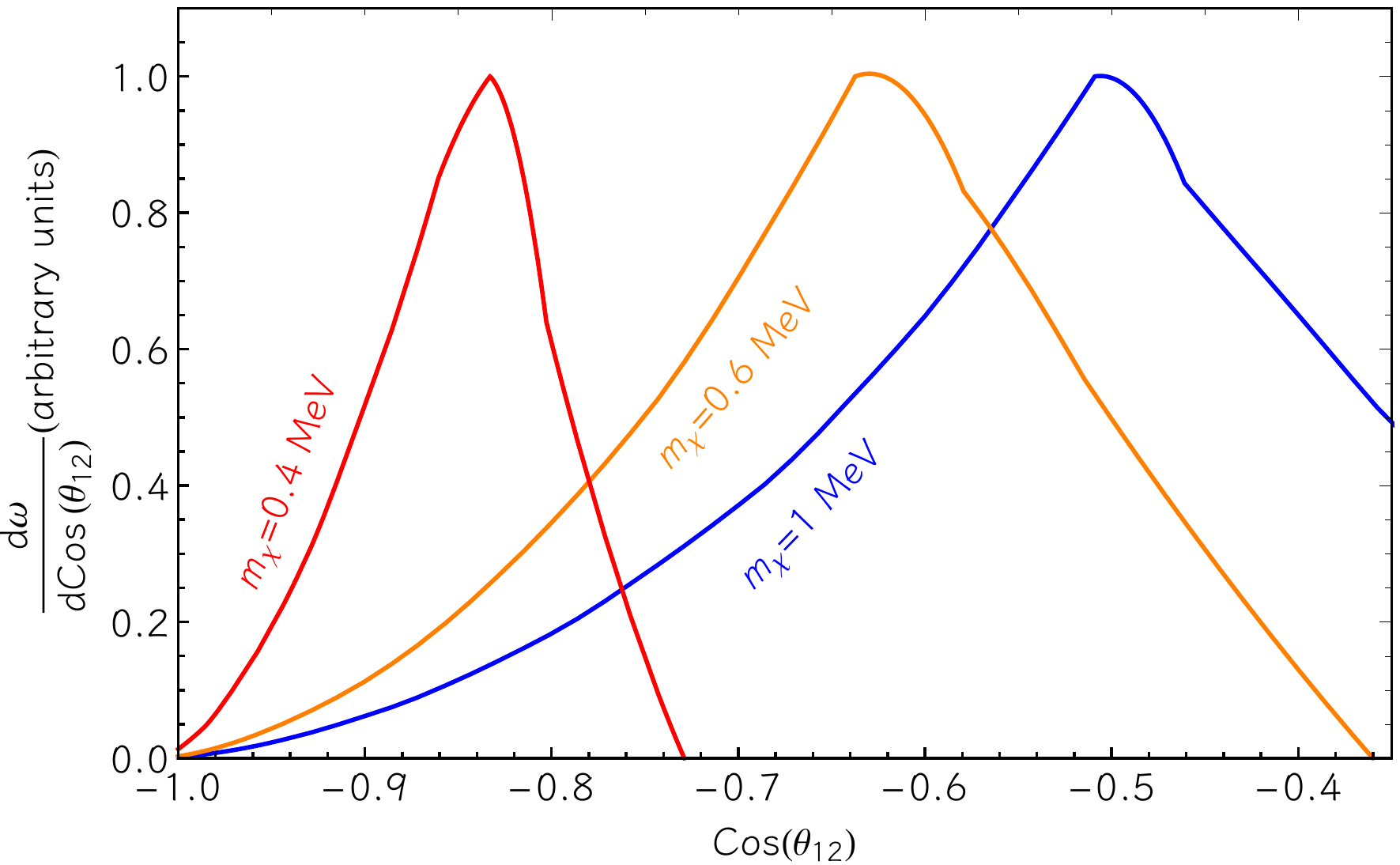}
\caption{Differential distribution of the energy released to the system as a function of the relative angle between the outgoing phonons and for few values of dark matter masses. The distributions have been normalized in order to have comparable amplitudes. The position of the peak depends on the mass of the dark matter candidate.} \label{fig:angulardist}
\end{figure}

As for the two-phonon emission, expression~\eqref{eq:gamma} allows to  compute the distribution of the energy deposited in the system, $\omega$, as a function of the angle between the two outgoing phonons. In Figure~\ref{fig:angulardist} we show it for different values of the dark matter mass. As one can see, this confirms that the lighter the dark matter the more back-to-back the outgoing phonons will be. Importantly, the position of the peak could thus provide a measurement of the dark matter mass.

\section{Conclusion}

In this work we studied $^4$He as a possible detector of dark matter with mass down to few keVs.  From the comparison with what obtained with time honored standard techniques, especially in the low dark matter mass region, we show that the EFT description of the phonon dynamics is impressively successful. Our final results rely only on the equation of state of superfluid helium, and are found to agree with what obtained with methods fitting data on neutron scattering.  
Being formulated in a quantum field theory language, the EFT allows to describe the phonon's interactions in a transparent and simple way, which can borrow high energy physics expertise and be accessible to non-experts of the microscopic theory of superfluid $^4$He. 

From the EFT we gain an understanding of the mechanism behind the strong suppression of the two-phonon emission rate, proving that it is a consequence of a fine tuned cancellation between different effective 
operators in the limit of small exchanged momenta.

We proved that this cancellation is only effective if the in-medium potential for the dark matter is precisely proportional to the $^4$He number density. Consequently, if the dark matter couples to the $^4$He via a different operator, the helium detector could be much more efficient at constraining the dark matter mass and couplings. In fact, while for larger masses, $m_{\chi} \gtrsim 1$~MeV, there are experiments and proposals competitive with the one analyzed here (see for example~\cite{Essig:2011nj,Essig:2016crl,Essig:2019kfe} or, for even larger masses, \cite{Agnes:2018ves}), for lower masses the present setup may provide the strongest constraints, without relying on any dark matter-electron coupling.

We also showed that the main contribution to the total rate of events comes from the phonons, which are exactly the degrees of freedom incorporated in EFT description.
Moreover, their angular distributions could be used to extract  information about the dark matter mass. This is true for the emission of  both one and two phonons.

Given the previous considerations, the use of $^4$He might play a central role in the future searches for sub-GeV dark matter, with the EFT providing a new theoretical tool, with a vast spectrum of possible applications.

\vspace{1em}

\begin{acknowledgments}
We are grateful to F.~Acanfora, G.~Cavoto, S.~De Cecco, S.~Knapen, T.~Lin, R.~Rattazzi, K.~Schutz, S.~Sun and K.~Zurek for useful discussions and comments. A.C. and A.E. would also like to thank CERN for the hospitality during the final stages of this project. A.E. is supported by the Swiss National Science Foundation under contract 200020-169696 and through the National Center of Competence in Research SwissMAP. A.C. is supported by grants FPA2014-57816-P, PROMETEOII/2014/050 and SEV-2014-0398, as well as by the EU projects H2020-MSCA-RISE-2015 and H2020-MSCA-ITN- 2015//674896-ELUSIVES.
\end{acknowledgments}

\bibliographystyle{apsrev4-1}
\bibliography{biblio}

\end{document}